\documentclass[a4paper,fleqn]{cas-dc}
\graphicspath{ {./figures/} }

\usepackage[authoryear,longnamesfirst]{natbib}
\usepackage{amsmath,amsbsy,amsmath,amsthm,amssymb,amsfonts}
\usepackage{graphicx,xcolor}
\usepackage{booktabs}
\usepackage{bm}
\usepackage{graphicx}
\usepackage{algorithmic,float}
\usepackage{array}
\usepackage[caption=false,font=normalsize,labelfont=sf,textfont=sf]{subfig}
\usepackage{textcomp}
\usepackage{stfloats}
\usepackage{url}
\usepackage{verbatim}
\usepackage{graphicx}
\usepackage[ruled,vlined, linesnumbered]{algorithm2e}
\usepackage{booktabs}
\usepackage{orcidlink}
\usepackage{enumerate}
\usepackage{epsfig}
\usepackage{rotating}
\usepackage{csvsimple}
\usepackage{lscape}
\usepackage{adjustbox}
\usepackage{titletoc}
\usepackage{longtable}
\usepackage{tabularx}
\usepackage{xtab}
\usepackage{afterpage}
\usepackage[T1]{fontenc}
\usepackage{mwe}  
\hypersetup{
	pdfsubject={Sequential asset ranking in nonstationary time series},
	pdfkeywords={online learning, prediction with expert advice, learning to rank, transfer learning, radial basis function networks, multivariate regression shrinkage},
	pdfauthor={Gabriel Borrageiro},
	pdftitle={Sequential asset ranking in nonstationary time series},
}

\def\tsc#1{\csdef{#1}{\textsc{\lowercase{#1}}\xspace}}
\tsc{WGM}
\tsc{QE}

\theoremstyle{definition}

\newdefinition{rmk}{Remark}
\newproof{pf}{Proof}
\newproof{pot}{Proof of Theorem \ref{thm}}

\begin{document}
	\let\WriteBookmarks\relax
	\def\floatpagepagefraction{1}
	\def\textpagefraction{.001}
	
	\shorttitle{Sequential asset ranking in nonstationary time series}    
	\shortauthors{Borrageiro et al.}  
	
	\title{Sequential asset ranking in nonstationary time series}  
	
	\author[1]{Gabriel Borrageiro}[orcid=0000-0002-0063-7103]
	\cormark[1]
	\ead{gabriel.borrageiro.20@ucl.ac.uk}
	
	\affiliation[1]{organization={University College London},
		addressline={Gower Street}, 
		city={London},
		postcode={WC1E 6BT}, 
		country={United Kingdom}}
	
	\author[1]{Nick Firoozye}[orcid=0000-0002-6460-0406]
	\author[1]{Paolo Barucca}[orcid=0000-0003-4588-667X]
	
	\cortext[1]{Corresponding author}
	
	\begin{abstract}
		We create a ranking algorithm, the naive Bayes asset ranker. Our algorithm computes the posterior probability that individual assets will be ranked higher than other portfolio constituents. Unlike earlier algorithms, such as the weighted majority, our algorithm allows poor-performing experts to have increased weight when they start performing well. We outperform the long-only holding of the S\&P 500 index and a regress-then-rank baseline.
	\end{abstract}
	
	
	\maketitle
	
	\section{Introduction}
	Our particular modelling interest is in financial time series, which are typically nonstationary. Nonstationarity implies statistical distributions that adapt over time and violates the independent and identically distributed (iid) random variables assumption of most regression and classification models. We require approaches that adopt sequential optimisation methods, preferably methods that make little or no assumptions about the data-generating process. The main result of this paper is our novel ranking algorithm, the naive Bayes asset ranker, which we use to select subsets of assets to trade from the S\&P 500 index in either a long-only or a long/short (cross-sectional momentum) capacity. We achieve higher risk-adjusted and total returns than a strategy that would hold the long-only S\&P 500 index with hindsight, despite the index appreciating by 205\% during the test period. We also outperform a regress-then-rank baseline, a sequentially fitted curds and whey \citep{curds_and_whey} multivariate regression model.
	
	\section{The naive Bayes asset ranker} \label{sec:nbar}
	Our ranking algorithm is the naive Bayes asset ranker (nbar). The nbar sequentially ranks a set of experts, estimating the one-step-ahead posterior probability that individual experts will be ranked higher than the remaining experts. In the context of the experiment described in section \ref{sec:research_experiment}, each expert is a forecasted return for an individual portfolio constituent of the S\&P 500. The forecasted returns come from the curds and whey (caw) multivariate regression model, which utilises feature representation transfer from the constituent S\&P 500 returns to radial basis function networks (rbfnets) \citep{MoodyDarken} whose k-means++ \citep{ArthurVassilvitskii2007} clusters form hidden units. Assume that the nbar is presented with a set of $q$ forecasts. The goal is to select a subset of experts $1 \leq k \leq q$ such that the reward of the $k$ experts is expected to be the highest; this is achieved by estimating the sequential posterior probability that expert $j \in 1, ..., q$ is ranked higher than each of the remaining $q-1$ experts. This posterior probability is computed with exponential decay, allowing experts who performed poorly and now perform well to be selected with greater weight than previously. 
	
	\section{The research experiment} \label{sec:research_experiment}
	Our research experiment aims to assess the benefits of sequentially optimised ranking algorithms to select subsets or portfolios of financial assets to hold in either a long-only or long/short (cross-sectional momentum) capacity. More concretely, we experiment with the constituents of the S\&P 500 index. We use the nbar as our portfolio selection algorithm. In order to assess the benefits of our ranking meta-model, we adopt a baseline, the long-only holding of the S\&P 500 constituents with equal weighting. This baseline replicates a passive, index-tracking investment strategy. 
	
	\subsection{The S\&P 500 dataset}
	We conduct this research experiment using the daily closing constituent prices for the S\&P 500 index, which we extract from Refinitiv. Due to their relatively new trade history, some time series have little data. Therefore, we select a subset of the S\&P 500 index, where each constituent contains a trade count greater than or equal to the 25'th percentile of trade counts; this leaves us with a subset of 378 Refinitiv information codes (rics). The dataset begins on 2001-01-26 and ends on 2022-03-25, 5326 days.
	
	\subsection{Experiment design} \label{chapter_5_sec_experiment design}
	We use the first 25\% of the data as a training set and the remaining data as a test set. The caw and nbar models are initialised and fitted in the training set. These models are also sequentially optimised without forward-looking bias in the test set. Once the training data are assigned to their nearest cluster centres, the cluster-conditional covariance matrices and their inverses are estimated. Cluster centres with few training data vectors assigned to them are regularised to a diagonal variance prior. Thus, we are adopting a Bayesian maximum a posteriori procedure here.
	
	We use the forecasts of the caw model as the basis for taking risk in a subset of constituents in the S\&P 500 index. Specifically, the long-only caw model buys the expected top five per cent of performing assets with equal weight. The long/short caw model works similarly, except that it includes the short-selling of the bottom five per cent of most negative forecasts. A second forecaster we consider is the nbar algorithm, applied to the one-step-ahead forecasts of the caw model. The nbar selects portfolio constituents with weights determined by the posterior ranking probabilities. We must also consider execution costs. We force the caw and nbar models to trade as price takers, meaning that the models incur a cost equal to half the bid/ask spread times the change in absolute position. Furthermore, as these data are sampled daily, any portfolio rebalancing is applied at most once a day, at the close of trading circa 4 pm EST. 
	
	\subsection{Results}
\begin{figure}[htpb]
		\centering
		\includegraphics[width=1.\columnwidth]{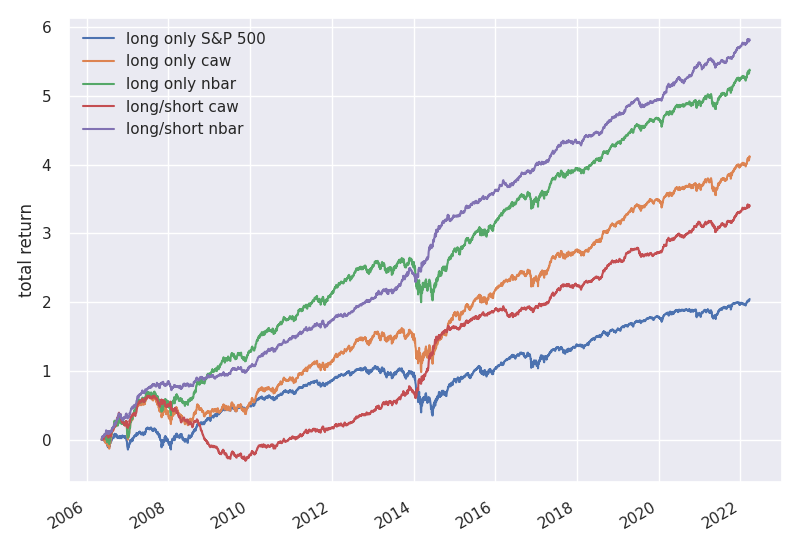}
		\caption{Total return by each model in the test set where the maximum selection percentile is set to 5\% of the total number of portfolio constituents. The naive Bayes asset ranker performs best, particularly the cross-sectional momentum version.}
		\label{fig:sp500_pnl}
	\end{figure}

	The passive index tracking baseline purchases each constituent with equal weighting at $t=0$ and holds them till the end of the experiment. This strategy pays transaction costs once and therefore has the least fees, as shown in table \ref{tab:sp500_transaction_costs}. Table \ref{tab:sp500_pnl} and figure \ref{fig:sp500_pnl} also show that the cumulative returns generated by this strategy are 205\%, the compound annual growth rate (cagr) is 7.3\% and the risk-adjusted annualised Sharpe ratio (sr) is a little under $0.8$. Assuming normally distributed returns, the Sharpe ratio implies a probability of positive annual returns of 71\%. The largest peak-to-trough drawdown for the strategy is just under 72\%, and the total return to maximum drawdown is around 2.9. Finally, by simply holding the index, the percentage of days with positive returns is 55\%. 
	
	The same performance metrics are available for the caw and nbar models. Both long-only and long/short caw and nbar models outperform the passive index tracking baseline, with the long/short models showing higher risk-adjusted performance measures indicated by the Sharpe ratios. The nbar performs best, with the long/short nbar showing the highest total and risk-adjusted returns. Table \ref{tab:sp500_transaction_costs} shows that despite the caw and nbar models being actively managed strategies that rebalance the portfolios daily, only the caw models show high transaction costs. The nbar models rebalance less often and do a better job of picking portfolio constituents. 
	
	\begin{table}[htpb]
		\centering
		\caption{Relative transaction costs incurred by each model in the test set. A buy-and-hold strategy on the S\&P 500 achieves the lowest transaction costs. However, from the perspective of a more active portfolio management standpoint, our ranking algorithm incurs far lower transaction costs than the regress-then-rank baseline.}
		\label{tab:sp500_transaction_costs}
		\begin{tabular}{lr}
			\toprule
			{} &   transaction costs \\
			\midrule
			long S\&P 500 & \textbf{-0.003} \\
			long caw     & -0.933 \\
			long/short caw    & -1.966 \\
			long nbar    & -0.050 \\
			long/short nbar   & -0.104 \\
			\bottomrule
		\end{tabular}
	\end{table}
	
	\begin{table}[htpb]
		\centering
		\caption{Summary returns statistics are shown in relation to the experiment, shown visually in figure \ref{fig:sp500_pnl}. The cross-sectional momentum naive Bayes asset ranker has the highest total and risk-adjusted returns.}
		\label{tab:sp500_pnl}
		\begin{tabularx}{1\columnwidth}{lXXXXXX}
			\toprule
			{} &  long S\&P 500 &  long caw &  long nbar &  long short caw &  long short nbar \\
			\midrule
			mean               &     0.0005 &     0.001 &      0.0013 &      0.0009 &      0.0015 \\
			std                &    0.012 &     0.016 &      0.016 &      0.010 &      0.010 \\
			total ret                &       2.047 &     4.113 &      5.372 &      3.397 &      \textbf{5.806} \\
			cagr               &       0.073 &     0.108 &      0.124 &      0.098 &      0.128 \\
			sr                 &       0.798 &     1.243 &      1.636 &      1.624 &      \textbf{2.879} \\
			$pr(\text{ann. ret} > 0)$ &       0.71 &     0.816 &      0.895 &      0.893 &      0.994 \\
			max dd             &       0.717 &     0.64 &      0.646 &      0.942 &      0.202 \\
			total ret / max dd       &       2.853 &     6.423 &      8.311 &      3.607 &      28.7 \\
			win ratio \%       &       0.549 &     0.553 &      0.563 &      0.547 &      0.575 \\
			\bottomrule
		\end{tabularx}
	\end{table}

	\section{Discussion} \label{chapter_5_sec_discussion}
	The shortcomings of regression models over classification ones in financial time series modelling are well-understood. \cite{Satchell_Timmermann_1995} show that regression models that typically minimise prediction mean-square error (mse) obtain worse performance than a random-walk model when forecasting daily foreign exchange (fx) returns. Furthermore,  they show that the probability of correctly predicting the sign of the change in daily fx rates is higher for the regression models than the random-walk baseline, even though the mse of the regression models exceeds that of the random-walk model. They conclude that mse is only sometimes an appropriate performance measure for evaluating predictive performance. More recently, \cite{pmlr-v55-amjad16} find that classical time series regression algorithms, such as arima models, have poor performance when forecasting Bitcoin returns. However, they find that the probability distribution of the sign of future price changes is adequately approximated from finite data, specifically classification algorithms that estimate this conditional probability distribution. 
	
	\section{Conclusions}
	We extend the research into cross-sectional momentum trading strategies. Our main result is our novel ranking algorithm, the naive Bayes asset ranker (nbar), which we use to select subsets of assets to trade from the S\&P 500 index. We perform feature representation transfer from radial basis function networks to a curds and whey (caw) multivariate regression model that takes advantage of the correlations between the response variables to improve predictive accuracy. The nbar ranks this regression output by forecasting the one-step-ahead sequential posterior probability that individual assets will be ranked higher than other portfolio constituents. Earlier algorithms, such as the weighted majority, deal with nonstationarity by ensuring the weights assigned to each expert never dip below a minimum threshold without ever increasing weights again. Our ranking algorithm allows experts who previously performed poorly to have increased weights when they start performing well. Our algorithm outperforms a strategy that would hold the long-only S\&P 500 index with hindsight, despite the index appreciating by 205\% during the test period. It also outperforms a regress-then-rank baseline, the caw model.

	\bibliographystyle{apalike}
	\bibliography{main}

\begin{thebibliography}{}

\bibitem[Amjad and Shah, 2017]{pmlr-v55-amjad16}
Amjad, M. and Shah, D. (2017).
\newblock Trading bitcoin and online time series prediction.
\newblock In {\em Proceedings of the Time Series Workshop at NIPS 2016},
  volume~55, pages 1--15, Barcelona, Spain. PMLR.

\bibitem[Arthur and Vassilvitskii, 2007]{ArthurVassilvitskii2007}
Arthur, D. and Vassilvitskii, S. (2007).
\newblock K-means++: The advantages of careful seeding.
\newblock In {\em Proceedings of the Eighteenth Annual ACM-SIAM Symposium on
  Discrete Algorithms}, page 1027–1035, New Orleans, Louisiana. Society for
  Industrial and Applied Mathematics.

\bibitem[Bishop, 1995]{BishopChristopherM1995Nnfp}
Bishop, C.~M. (1995).
\newblock {\em Neural networks for pattern recognition}.
\newblock Oxford University Press, Oxford.

\bibitem[Borrageiro et~al., 2021]{Borrageiro_Online_RBFNets}
Borrageiro, G., Firoozye, N., and Barucca, P. (2021).
\newblock Online learning with radial basis function networks.

\bibitem[Borrageiro et~al., 2022a]{Borrageiro_RL_crypto}
Borrageiro, G., Firoozye, N., and Barucca, P. (2022a).
\newblock The recurrent reinforcement learning crypto agent.
\newblock {\em {IEEE} Access}, 10:38590--38599.

\bibitem[Borrageiro et~al., 2022b]{Borrageiro_RL_FX}
Borrageiro, G., Firoozye, N., and Barucca, P. (2022b).
\newblock Reinforcement learning for systematic {FX} trading.
\newblock {\em {IEEE} Access}, 10:5024--5036.

\bibitem[Breiman and Friedman, 1997]{curds_and_whey}
Breiman, L. and Friedman, J.~H. (1997).
\newblock Predicting multivariate responses in multiple linear regression.
\newblock {\em Journal of the Royal Statistical Society: Series B (Statistical
  Methodology)}, 59(1):3--54.

\bibitem[Cesa-Bianchi and Lugosi, 2006]{Cesa-Bianchi_Lugosi_PLG}
Cesa-Bianchi, N. and Lugosi, G. (2006).
\newblock {\em Prediction, Learning, and Games}.
\newblock Cambridge University Press, USA.

\bibitem[Cover, 1991]{Cover_1991}
Cover, T.~M. (1991).
\newblock Universal portfolios.
\newblock {\em Mathematical Finance}, 1(1):1--29.

\bibitem[Cover and Ordentlich, 1996]{Cover_Ordentlich_1996}
Cover, T.~M. and Ordentlich, E. (1996).
\newblock Universal portfolios with side information.
\newblock {\em IEEE Transactions on Information Theory}, 42(2):348--363.

\bibitem[Flach and Matsubara, 2007]{Flach_Matsubara_2007}
Flach, P. and Matsubara, E.~T. (2007).
\newblock {\em A Simple Lexicographic Ranker and Probability Estimator}.
\newblock Springer Berlin Heidelberg, Berlin, Heidelberg.

\bibitem[Granger and Newbold, 1974]{GrangerNewbold1974}
Granger, C.~W. and Newbold, P. (1974).
\newblock Spurious regressions in econometrics.
\newblock {\em Journal of econometrics}, 2:111--120.

\bibitem[Gu et~al., 2020]{Gu_2020}
Gu, S., Kelly, B., and Xiu, D. (2020).
\newblock {Empirical Asset Pricing via Machine Learning}.
\newblock {\em The Review of Financial Studies}, 33(5):2223--2273.

\bibitem[Helmbold et~al., 1998]{Helmbold_portfolio_selection_1998}
Helmbold, D.~P., Schapire, R.~E., Singer, Y., and Warmuth, M.~K. (1998).
\newblock On-line portfolio selection using multiplicative updates.
\newblock {\em Mathematical Finance}, 8(4):325--347.

\bibitem[Jaeger, 2002]{Jaeger2002AdaptiveNS}
Jaeger, H. (2002).
\newblock Adaptive nonlinear system identification with echo state networks.
\newblock In {\em Advances in neural information processing systems},
  volume~15, pages 1000--1008, Vancouver, British Columbia. Advances in neural
  information processing systems.

\bibitem[Jegadeesh and Titman, 1993]{Jegadeesh1993}
Jegadeesh, N. and Titman, S. (1993).
\newblock Returns to buying winners and selling losers: Implications for stock
  market efficiency.
\newblock {\em The Journal of Finance}, 48:65--91.

\bibitem[Joao, 2012]{Gama2012}
Joao, G. (2012).
\newblock A survey on learning from data streams: current and future trends.
\newblock {\em Progress in AI}, 1(1):45--55.

\bibitem[Kivinen and Warmuth, 1995]{Kivinen_Warmuth_1995}
Kivinen, J. and Warmuth, M.~K. (1995).
\newblock Additive versus exponentiated gradient updates for linear prediction.
\newblock In {\em Proceedings of the Twenty-Seventh Annual ACM Symposium on
  Theory of Computing}, page 209–218, New York, NY, USA. ACM.

\bibitem[Krawczyk and Wozniak, 2015]{KrawczykWozniak2015}
Krawczyk, B. and Wozniak, M. (2015).
\newblock Weighted naïve bayes classifier with forgetting for drifting data
  streams.
\newblock In {\em 2015 IEEE International Conference on Systems, Man, and
  Cybernetics}, pages 2147--2152, Hong Kong. IEEE.

\bibitem[Littlestone and Warmuth, 1994]{Littlestone1994TheWM}
Littlestone, N. and Warmuth, M.~K. (1994).
\newblock The weighted majority algorithm.
\newblock {\em Information and Computation}, 108:212--261.

\bibitem[Lloyd, 1982]{LloydS1982Lsqi}
Lloyd, S. (1982).
\newblock Least squares quantization in pcm.
\newblock {\em IEEE transactions on information theory}, 28(2):129--137.

\bibitem[Markowitz and Cootner, 1965]{MarkowitzHarry1965TRCo}
Markowitz, H.~M. and Cootner, P.~H. (1965).
\newblock The random character of stock market prices.
\newblock {\em Journal of the American Statistical Association},
  60(309):381--381.

\bibitem[Merton, 1976]{Merton76}
Merton, R.~C. (1976).
\newblock Option pricing when underlying stock returns are discontinuous.
\newblock {\em Journal of Financial Economics}, 3(1):125 -- 144.

\bibitem[Moody and Darken, 1989]{MoodyDarken}
Moody, J. and Darken, C.~J. (1989).
\newblock Fast learning in networks of locally-tuned processing units.
\newblock {\em Neural computation}, 1:281--294.

\bibitem[Nakamura and Small, 2007]{NAKAMURA2007599}
Nakamura, T. and Small, M. (2007).
\newblock Tests of the random walk hypothesis for financial data.
\newblock {\em Physica A: Statistical Mechanics and its Applications},
  377(2):599--615.

\bibitem[Pan and Yang, 2010]{Pan2010ASO}
Pan, S.~J. and Yang, Q. (2010).
\newblock A survey on transfer learning.
\newblock {\em IEEE Transactions on Knowledge and Data Engineering},
  22:1345--1359.

\bibitem[Poh et~al., 2021]{Poh2021}
Poh, D., Lim, B., Zohren, S., and Roberts, S. (2021).
\newblock Building cross-sectional systematic strategies by learning to rank.
\newblock {\em The Journal of Financial Data Science}, 3(2):70--86.

\bibitem[Poh et~al., 2022]{Poh2022}
Poh, D., Lim, B., Zohren, S., and Roberts, S. (2022).
\newblock Enhancing cross-sectional currency strategies by context-aware
  learning to rank with self-attention.
\newblock {\em The Journal of Financial Data Science}, 4(3):89--107.

\bibitem[Provost and Fawcett, 1997]{Provost_Fawcett_1997}
Provost, F. and Fawcett, T. (1997).
\newblock Analysis and visualization of classifier performance: Comparison
  under imprecise class and cost distributions.
\newblock In {\em Proceedings of the Third International Conference on
  Knowledge Discovery and Data Mining}, page 43–48, Newport Beach, CA. AAAI
  Press.

\bibitem[Rakhlin and Sridharan, 2014]{Rakhlin_Sridharan_2014}
Rakhlin, A. and Sridharan, K. (2014).
\newblock Statistical learning theory and sequential prediction.
\newblock Technical report, MIT.
\newblock STAT928.

\bibitem[Said and Dickey, 1984]{SaidDickey1984}
Said, S.~E. and Dickey, D.~A. (1984).
\newblock {Testing for unit roots in autoregressive-moving average models of
  unknown order}.
\newblock {\em Biometrika}, 71(3):599--607.

\bibitem[Satchell and Timmermann, 1995]{Satchell_Timmermann_1995}
Satchell, S. and Timmermann, A. (1995).
\newblock An assessment of the economic value of non-linear foreign exchange
  rate forecasts.
\newblock {\em Journal of Forecasting}, 14(6):477--497.

\bibitem[Singer, 1998]{Singer98switchingportfolios}
Singer, Y. (1998).
\newblock Switching portfolios.
\newblock In {\em International Journal of Neural Systems}, pages 488--495,
  Burlington, Massachusetts. Morgan Kaufmann.

\bibitem[Snedecor and Cochran, 1989]{snedecor1989statistical}
Snedecor, G.~W. and Cochran, W.~G. (1989).
\newblock {\em Statistical Methods, eight edition}, volume 1191.
\newblock Iowa state University press, Ames, Iowa.

\bibitem[Sutton and Barto, 2018]{SuttonBarto2018}
Sutton, R.~S. and Barto, A.~G. (2018).
\newblock {\em Reinforcement Learning: An Introduction}.
\newblock A Bradford Book, Cambridge, MA, USA.

\bibitem[Yang et~al., 2020]{yang_zhang_dai_pan_2020}
Yang, Q., Zhang, Y., Dai, W., and Pan, S.~J. (2020).
\newblock {\em Transfer Learning}.
\newblock Cambridge University Press, Cambridge.

\bibitem[Zhang and Su, 2004]{Zhang_Su_NBR_2004}
Zhang, H. and Su, J. (2004).
\newblock Naive bayesian classifiers for ranking.
\newblock In Boulicaut, J.-F., Esposito, F., Giannotti, F., and Pedreschi, D.,
  editors, {\em Machine Learning: ECML 2004}, pages 501--512, Berlin,
  Heidelberg. Springer Berlin Heidelberg.

\end{thebibliography}
	
\end{document}